\documentstyle[12pt]{article}
\begin{document}
\begin{titlepage}
{\hskip 11cm} 
\vspace*{2.7cm}
\begin{center}
{\Large The Spin Symmetry of Heavy Baryons 
in the Framework of the B.S. Equation\\}
\vspace*{0.4cm}
{{J.Y. Cui $^a$ $^ b$  H.Y. Jin $^a$ and J.M. Wu $^a$}\\
  {$^a$ \small Institute of High Energy Physics}\\
  {\small Beijing, 100039, P.R. China}\\
  {$^b$ \small Department of Physics, Henan Normal University}\\
  {\small Xinxiang, 453002, P.R. China}}
\end{center}
\date{}
\begin{center}
\begin{minipage}{120mm}
\vspace*{1.5cm}
\begin{center}{\bf Abstract}\end{center}
{We construct the B.S. equation for baryons.  The most general form of 
the B.S. wave function is given. In the heavy quark limit we show
clearly  that the spin symmetry exists in  heavy baryon states.\\
{\bf PACS: 14.20, 11.10.S. }}
\end{minipage}
\end{center}
\vskip 1in
\end{titlepage}
\newpage
1.{\it Introduction}{~~~} It's known the physics of heavy hadrons is
important. The weak decays of heavy quark are employed
for the tests of the standard model and the measurements of its parameter.
For example, they offer the most direct way to determine the weak mixing
angles and to test the unitarity of Kobayashi-Maskawa matrix. In this paper
we study baryons composed of a heavy quark and other light constituents in
the framework of B.S. equation [1]. 
\par
The typical four-momentum of the light constituent is small compared with the
heavy quark mass, and the typical momentum exchanged between the heavy quark
and light constituents are order of $\Lambda_{QCD}$, therefore, the strong
interaction physics of such system is mainly nonperturbative. In this
situation, it's suitable to carry out the systematic expansions in $1/M_Q$
based on QCD, then the so called heavy quark effective theory (HQET) is
obtained [2-8]. In the leading order of HQET ($M\rightarrow \infty$), the strong
interactions of the heavy quark are independent of its mass and spin.
Consequently, for $N$ heavy quarks, there will be $SU(2N)$ spin-flavor
symmetry in the leading order  of HQET. This symmetry directly leads to many
predictions. As one of the important results, this symmetry is responsible
for some mass degeneracy in hadron spectrum. 
\par
In fact, the the systematic expansions in $1/M_Q$ can also be carried out
in the B.S. formalism. For general case, the B.S. wave function of hadrons
have many components, and the numerical calculation of the relevant integral
equation is very difficult. In the case of mesons, it has been shown that the 
number of components of the general relativistic covariant B.S. wave function 
is reduced greatly in the leading order of $1/M_Q$ expansion, and the
spin-flavor symmetry is shown clearly [9]. In this paper, we extend this 
method to the system
of baryons. The paper is arranged as follows. In section two, we construct
the B.S. equation for baryons, and we give the general relativistic B.S.
equation. In section three, we study the spin symmetry in heavy quark limit
in the framework of B.S. equation. In section three results and discussion
are presented.

2.{\it The B.S. equation and the B.S. wave function }{~~~} Let
$\psi_1(x_1)$, $\psi_2(x_2)$ and $\psi_3(x_3)$ be the quark fields at points
$x_1$, $x_2$ and $x_3$, $|B\rangle$ the baryon state with momentum $P$
and mass $M$. The the B.S. wave function of the baryons is defined as
$$\begin{array}{c}
\chi_P(x_1,x_2,x_3)=\langle |T\psi_1(x_1)\psi_2(x_2)\psi_3(x_3)
|B\rangle
\end{array}
\eqno(1)$$
In order to separate the center of mass coordinates and the internal relative
coordinates, we use the following variables
$$\begin{array}{l}
\displaystyle X=\frac{1}{m_1+m_2+m_3}(m_1x_1+m_2x_2+m_3x_3)
\end{array}
\eqno(2)$$
$$\begin{array}{c}
\displaystyle x=x_2-x_3, \hskip 0.4in 
x^\prime=x_1-\frac{1}{m_2+m_3}(m_2x_2+m_3x_3)
\end{array}
\eqno(3)$$
Then we have
$$\begin{array}{c}
\chi_P(x_1,x_2,x_3)=e^{-iP\!\cdot\! X}\chi_P(x,x^\prime)
\end{array}
\eqno(4)$$
In momentum space, we define the B.S. amplitude as
$$\begin{array}{c}
\displaystyle\chi(P,q,k)=\int d^4x d^4x^\prime 
e^{iq\!\cdot\!x}e^{ik\!\cdot\!x^\prime}\chi_P(x,x^\prime)
\end{array}
\eqno(5)$$
With a standard method, we obtain the B.S. equation for baryons in momentum
space
$$\begin{array}{l}
\displaystyle\chi(P,q,k)=S^{(1)}_F(p_1)S^{(2)}_F(p_2)S^{(3)}(p_3)\int
\frac{d^4q^\prime}{(2\pi)^4}\frac{d^4k^\prime}{(2\pi)^4}
G(P,q,q^\prime,k,k^\prime)\chi(p,q^\prime,k^\prime)
\end{array}
\eqno(6)$$
where $S_F(p_i)$'s are the propagators of the quark with momentum $p_i$, and
$G(P,q,q^\prime,k,k^\prime)$ is the kernel defined as all irreducible
three-particle graphs.
\par{~~~}Now we construct the the B.S. wave functions of baryons.
Generally, the B.S. wave functions of baryons are
complicated. The spin-$(j+\frac{1}{2})$ wave functions
can be written as the sum of the terms  
$u^{\mu_1\mu_2\ldots \mu_j}_\alpha
\left\{\Gamma_{\mu_1\mu_2\ldots \mu_j}\right\}_{\beta\gamma}$,
where $u^{\mu_1\mu_2\ldots \mu_j}_\alpha$ is a tensor spinor, and $\Gamma$ a tensor matrix
made of $P_\mu$, $q_\mu$ and $k_\mu$ and $\gamma$-matrices.
The tensor spinor $u^{\mu_1\mu_2\ldots \mu_j}$ is symmetric about the
Lorentz indices $\mu_1$, $\mu_2$, $\ldots$, $\mu_j$, 
and satisfies the subsidiary conditions
$$\begin{array}{l}
(1-v\!\!\!/~)u^{\mu_1\mu_2\ldots \mu_j}=0, \hskip 0.4in
u_\mu^{\mu\mu_3,\ldots \mu_j}=0
\end{array}
\eqno(7)$$
$$\begin{array}{l}
v_{\mu_1} u^{\mu_1\mu_2\ldots \mu_j}=0, \hskip 0.4in
\gamma_{\mu_1} u^{\mu_1\mu_2\ldots \mu_j}=0.
\end{array}
\eqno(8)$$
where $v=P/M$. \\
For convenience,following we define some tensor matrices
$$\begin{array}{l}
\Gamma =f_1 + f_2 q\!\!\!/_\bot + f_3 k\!\!\!/_\bot
+ f_4 q\!\!\!/_\bot k\!\!\!/_\bot + v\!\!\!/~
(f_5 + f_6 q\!\!\!/_\bot + f_7 k\!\!\!/_\bot
+ f_8 q\!\!\!/_\bot k\!\!\!/_\bot)
\end{array}
\eqno(9)$$
$$\begin{array}{l}
\Gamma^\mu=q_\bot^\mu\Gamma_1 + k_\bot^\mu\Gamma_2
+ \gamma_\bot^\mu\Gamma_3
\end{array}
\eqno(10)$$
$$\begin{array}{l}
\Gamma^{\mu\nu}=q_\bot^\mu \Gamma_{1}^\nu
+ k_\bot^\mu \Gamma_{2}^\nu
+ \gamma_\bot^\mu \Gamma_{3}^\nu
\end{array}
\eqno(11)$$
$$\begin{array}{l}
\Gamma^{\mu\nu\lambda}=q_\bot^\mu \Gamma_{1}^{\nu\lambda}
+ k_\bot^\mu \Gamma_{2}^{\nu\lambda}
+ \gamma_\bot^\mu \Gamma_{3}^{\nu\lambda}
\end{array}
\eqno(12)$$
\hskip 2.8in \ldots \ldots\\
where $q_\bot$, $k_\bot$  $\gamma_\bot^\mu$ (and $\sigma_\bot^{\mu\nu}$ and
$g_\bot^{\mu\nu}$ to be used later) 
are transverse quantities defined as
$$\begin{array}{l}
q_\bot^\mu=q^\mu - v\!\cdot\!q v^\mu, ~~~~~~
k_\bot^\mu=q^\mu - v\!\cdot\!k v^\mu
\end{array}
\eqno(13)$$
$$\begin{array}{l}
\gamma_\bot^\mu=\gamma^\mu - v\!\!\!/~ v^\mu, ~~~~~~
\sigma_\bot^{\mu\nu}=\frac{1}{2}[\gamma_\bot^\mu,\gamma_\bot^\nu]
\end{array}
\eqno(14)$$
$$\begin{array}{l}
g_\bot^{\mu\nu}=g^{\mu\nu}-v^\mu v^\nu
\end{array}
\eqno(15)$$
where $f_i$ (i=1,2,\ldots ,8) are scalar functions of $P\!\cdot\!q$, $P\!\cdot\!k$,
$q\!\cdot\!k$ \ldots ,
$\Gamma_i$ (or $\Gamma_i^\mu$, $\Gamma_i^{\mu\nu}$) have the same structure as
$\Gamma$ (or $\Gamma_\mu$, $\Gamma_{\mu\nu}$) but with independent scalar
functions(ISF). So $\Gamma^\mu$ has $24$ ISF and
$\Gamma^{\mu\nu}$ has $72$ ISF, and so on.
The B.S. wave function can be construct now. For example, the wave function 
of spin-$\frac{1}{2}$ baryons can be written as
$$\begin{array}{l}
\chi^{(\frac{1}{2})+}(p,q,k)=u\Gamma^{(\frac{1}{2})+}_1\gamma_5 C 
+ \gamma_5u\Gamma^{(\frac{1}{2})+}_2 C
+ \gamma_\bot^\mu u \Gamma^{(\frac{1}{2})+}_{1~\mu}\gamma_5 C\\
\hskip 0.7in
+ \gamma_5\gamma_\bot^\mu u \Gamma^{(\frac{1}{2})+}_{2~\mu} C
+\sigma_\bot^{\mu\nu}u\Gamma^{(\frac{1}{2})+}_{\mu\nu}\gamma_5 C\\
\end{array}
\eqno(16)$$
For spin-$\frac{3}{2}$ baryons, the B.S. wave function can be written as
$$\begin{array}{l}
\chi^{(\frac{3}{2})+}(P,q,k)=u^\mu \Gamma^{(\frac{3}{2})+}_{1~\mu} C
+ \gamma_5 u^\mu \Gamma^{(\frac{3}{2})+}_{2~\mu}\gamma_5 C
+\gamma_\bot^\nu u^\mu\Gamma^{(\frac{3}{2})+}_{1~\mu\nu}C\\
\hskip 0.7in
+\gamma_5\gamma_\bot^\nu u^\mu\Gamma^{(\frac{3}{2})+}_{2~\mu\nu}\gamma_5C
+\sigma_\bot^{\lambda\nu}u^{\mu}
\Gamma^{(\frac{3}{2})+}_{\mu\nu\lambda}C\\
\end{array}
\eqno(17)$$
where $C$ is the charge conjugation matrix defined as
$$\begin{array}{l}
C^T=-C, ~~~~~~~~ C^{-1}\gamma^\mu C=-\gamma^{\mu T}
\end{array}
\eqno(18)$$
For the convenience of the following discussion, we rewrite the wave function 
(17) according to their symmetric property as
$$\begin{array}{l}
\chi^{\frac{3}{2}+}(P,q,k)=u^\mu \Gamma^{(\frac{3}{2})+}_{1~\mu} C
+ \gamma_5 u^\mu \Gamma^{(\frac{3}{2})+}_{2~\mu} \gamma_5 C\\
\hskip 0.7in
+ (\gamma_\bot^\nu u^\mu - \gamma_\bot^\mu u^\nu)
\Gamma^{(\frac{3}{2})+}_{1~\mu\nu}C
+ (\gamma_\bot^\nu u^\mu + \gamma_\bot^\mu u^\nu)
\Gamma^{(\frac{3}{2})+}_{2~\mu\nu}C\\
\hskip 0.7in
+ \gamma_5 (\gamma_\bot^\nu u^\mu - \gamma_\bot^\mu u^\nu)
\Gamma^{(\frac{3}{2})+}_{3~\mu\nu}\gamma_5C
+ \gamma_5 (\gamma_\bot^\nu u^\mu + \gamma_\bot^\mu u^\nu)
\Gamma^{(\frac{3}{2})+}_{4~\mu\nu}\gamma_5C\\
\hskip 0.7in
+[\sigma_\bot^{\lambda\nu}u^{\mu} -\frac{1}{2}( g_\bot^{\lambda\mu}u^\nu 
- g_\bot^{\nu\mu}u^\lambda)]\Gamma^{(\frac{3}{2})+}_{\mu\nu\lambda}C\\
\end{array}
\eqno(19)$$
It should be noticed that, in the last term of Eq.(16), $\sigma_\bot
^{\mu\nu}$ is
anti-symmetry, so $\Gamma^{({1\over2})+}_{\mu\nu}$ has 32 ISF
instead of 72. Similarly, in Eq(19), $\Gamma^{({3\over2})+}_{1~\mu\nu}$
has 32 ISF, $\Gamma^{({3\over2})+}_{2~\mu\nu}$ has 40 ISF and so on. 
Therefore, wave functions (17) and (19) have the same number of ISF.
One may also note that all tensors in front of $\Gamma^{({3\over2})+}_{i~\mu\nu}(i=1,2,3)$ and
$\Gamma^{({3\over2})+}_{\mu\nu\lambda}$ in (19) vanish after contraction of any two Lorenz
indices. So we neglect the terms proportional $g^{\mu\nu}_\bot$ in Eq.(11) and
Eq.(12), which is also unnecessary in Eq.(17), since
$\sigma^{\lambda\nu}_\bot u^{\mu}g_{\bot\mu\nu}=-u^\lambda$. 
Similarly, the wave function of spin-$\frac{5}{2}$ baryon state is written as
$$\begin{array}{l}
\chi^{(\frac{5}{2})+}(P,q,k)=u^{\mu\nu}
\Gamma^{(\frac{5}{2})+}_{1~\mu\nu}\gamma_5C
+ \gamma_5u^{\mu\nu}\Gamma^{(\frac{5}{2})+}_{2~\mu\nu}C\\
\hskip 0.7in
+(\gamma_\bot^\mu u^{\nu\lambda}
-\gamma_\bot^\nu u^{\mu\lambda})
\Gamma^{(\frac{5}{2})+}_{1~\mu\nu\lambda}\gamma_5C

+(\gamma_\bot^\mu u^{\nu\lambda}
+\gamma_\bot^\nu u^{\mu\lambda}
+\gamma_\bot^\lambda u^{\mu\nu})
\Gamma^{(\frac{5}{2})+}_{2~\mu\nu\lambda}\gamma_5C\\
\hskip 0.7in

+\gamma_5(\gamma_\bot^\mu u^{\nu\lambda}
-\gamma_\bot^\nu u^{\mu\lambda})
\Gamma^{(\frac{5}{2})+}_{3~\mu\nu\lambda}C

+\gamma_5(\gamma_\bot^\mu u^{\nu\lambda}
+\gamma_\bot^\nu u^{\mu\lambda}
+\gamma_\bot^\lambda u^{\mu\nu})
\Gamma^{(\frac{5}{2})+}_{4~\mu\nu\lambda}C\\
\hskip 0.7in

+[\sigma_\bot^{\mu\nu}u^{\lambda\rho}
-\frac{1}{3}(g_\bot^{\mu\lambda}u^{\nu\rho}+g_\bot^{\mu\rho}u^{\nu\lambda})

+\frac{1}{3}(g_\bot^{\nu\lambda}u^{\mu\rho}+g_\bot^{\nu\rho}u^{\mu\lambda})]
\Gamma^{(\frac{5}{2})+}_{\mu\nu\lambda\rho}\gamma_5C
\end{array}
\eqno(20)$$
\par
If we adopt  that the tensor spinor $u$ is the positive eigenstate of
operator $v\!\!\!/$, the above B.S. wave equations have positive parity.
The wave function with negative parity can be constructed easily by
multiplying a $\gamma_5$. For example, the wave function of 
spin-$\frac{1}{2}$ baryon with negative parity can be written as
$$\begin{array}{l}
\chi^{(\frac{1}{2})-}(p,q,k)=u\Gamma_1^{(\frac{1}{2})-} C 
+ \gamma_5u\Gamma^{(\frac{1}{2})-}_2\gamma_5C
+ \gamma_\bot^\mu u \Gamma^{(\frac{1}{2})-}_{1~\mu} C\\
\hskip 0.7in
+ \gamma_5\gamma_\bot^\mu u \Gamma^{(\frac{1}{2})-}_{2~\mu} \gamma_5C
+\sigma_\bot^{\mu\nu}u\Gamma^{(\frac{1}{2})-}_{\mu\nu}C\\
\end{array}
\eqno(21)$$
and for spin-$\frac{3}{2}$ state, the  wave function with negative parity 
is written as
$$\begin{array}{l}
\chi^{(\frac{3}{2})-}(P,q,k)=u^\mu \Gamma_{1~\mu}^{(\frac{3}{2})-}\gamma_5 C
+ \gamma_5 u^\mu \Gamma^{(\frac{3}{2})-}_{2~\mu} C\\
\hskip 0.7in
+ (\gamma_\bot^\nu u^\mu - \gamma_\bot^\mu u^\nu)
\Gamma^{(\frac{3}{2})-}_{1~\mu\nu}\gamma_5C
+ (\gamma_\bot^\nu u^\mu + \gamma_\bot^\mu u^\nu)
\Gamma^{(\frac{3}{2})-}_{2~\mu\nu}\gamma_5C\\
\hskip 0.7in
+ \gamma_5 (\gamma_\bot^\nu u^\mu - \gamma_\bot^\mu u^\nu)
\Gamma^{(\frac{3}{2})-}_{3~\mu\nu}C
+ \gamma_5 (\gamma_\bot^\nu u^\mu + \gamma_\bot^\mu u^\nu)
\Gamma^{(\frac{3}{2})-}_{4~\mu\nu}C\\
\hskip 0.7in
+[\sigma_\bot^{\lambda\nu}u^{\mu} -\frac{1}{2}( g_\bot^{\lambda\mu}u^\nu 
- g_\bot^{\nu\mu}u^\lambda)]\Gamma^{(\frac{3}{2})-}_{\mu\nu\lambda}\gamma_5C\\
\end{array}
\eqno(22)$$
\par
3.{\it The Spin Symmetry of the Heavy baryons}{~~~~}
When $m_1$ goes to infinity, the momentum of the heavy quark can be written as
$$\begin{array}{l}
\displaystyle
p^\mu_1={m_1\over{m_1+m_2+m_3}}P^\mu+k^\mu=(m_1+E)v^\mu+k^\mu
\end{array}
\eqno(23)$$
where $E=M-(m_1+m_2+m_3)$.
The propagator of the heavy quark is then simplified to
$$\begin{array}{l}
\displaystyle 
S^{(1)}_F=\frac{1}{E+k\!\cdot\!v}\frac{1+v\!\!\!/}{2}
\end{array}
\eqno(24)$$
Then equation (6) is reduced to
$$\begin{array}{l}
\displaystyle\chi(P,q,k)=\frac{1}{E+k\!\cdot\!v}\frac{1+v\!\!\!/}{2}
S^{(2)}_F(p_2)S^{(3)}(p_3)\int
\frac{d^4q^\prime}{(2\pi)^4}\frac{d^4k^\prime}{(2\pi)^4}
G(P,q,q^\prime,k,k^\prime)\chi(p,q^\prime,k^\prime)
\end{array}
\eqno(25)$$
From equation above, we have
$$\begin{array}{l}
\displaystyle
\chi(P,q,k)=\frac{1+v\!\!\!/}{2}\chi(P,q,k)
\end{array}
\eqno(26)$$
Therefore, all the components of the wave function, which are the sum of the
forms ($\gamma_5u^{\mu_1\mu_2\ldots }\Gamma_{\mu_1\mu_2\ldots }$) or
($\gamma_\bot^{\mu_1}u^{\mu_2\mu_3\ldots }\Gamma_{\mu_1\mu_2\ldots }$), will 
be reduced to zero.\\
Since
$\displaystyle \frac{1+v\!\!\!/}{2}\gamma^\mu\frac{1+v\!\!\!/}{2}=v^\mu$, 
in general the kernel can be written as the form
$$\begin{array}{l}
G(P,q,q^\prime,k,k^\prime)=I\bigotimes\tilde{G}
\end{array}
\eqno(27)$$
where $I$ belongs to the heavy quark spin space, and $\tilde{G}$ belongs to
the spin space of the light quarks.
For spin-$\frac{1}{2}$ baryon with positive parity,
substituting the wave function (16) into equation (25), we obtain 
three decoupled equations
$$\begin{array}{l}
\displaystyle
\Gamma^{(\frac{1}{2})+}_1\gamma_5 C
=\frac{1}{E+k\!\cdot\!v}S^{(2)}_F S^{(3)}_F
\int\frac{d^4q^\prime}{(2\pi)^4}\frac{d^4k^\prime}{(2\pi)^4}
\tilde{G}\Gamma^{(\frac{1}{2})+}_1\gamma_5 C 
\end{array}
\eqno(28)$$
$$\begin{array}{l}
\displaystyle
\Gamma^{(\frac{1}{2})+}_{2~\mu} C = \frac{1}{E+k\!\cdot\!v}S^{(2)}_F S^{(3)}_F
\int\frac{d^4q^\prime}{(2\pi)^4}\frac{d^4k^\prime}{(2\pi)^4}\tilde{G}
\Gamma^{(\frac{1}{2})+}_{2~\mu} C 
\end{array}
\eqno(29)$$
$$\begin{array}{l}
\displaystyle
(\Gamma^{(\frac{1}{2})+}_{\mu\nu} - 
\Gamma^{(\frac{1}{2})+}_{\nu\mu})\gamma_5C =
\frac{1}{E+k\!\cdot\!v}S^{(2)}_FS^{(3)}_F
\int\frac{d^4q^\prime}{(2\pi)^4}\frac{d^4k^\prime}{(2\pi)^4}\tilde{G}
(\Gamma^{(\frac{1}{2})+}_{\mu\nu} - 
\Gamma^{(\frac{1}{2})+}_{\nu\mu})\gamma_5C 
\end{array}
\eqno(30)$$
Similarly, for spin-$\frac{3}{2}$ baryon with positive parity,
substituting the wave function (19) into equation (25), we can obtain
four decoupled equations. One notice that the tensors
$(\gamma_\bot^\mu u^\nu+\gamma_\bot^\nu u^\mu)$ and $u^{\mu\nu}$ have the same
symmetry charactor and so do the tensors
$[\sigma_\bot^{\lambda\nu}u^\mu-\frac{1}{2}(g_\bot^{\lambda\mu}
u^\nu-g_\bot^{\nu\mu}u^\lambda)]$
and $(\gamma_\bot^\lambda u^{\mu\nu}-\gamma_\bot^\nu u^{\mu\lambda})$.
It's easy to see that the equation about the component
$u^\mu\Gamma^{(\frac{3}{2})+}_{1~\mu} C$ is just the same as equation (29), 
and the equation about the component
$\gamma_5(\gamma_\bot^\mu u^\nu -\gamma_\bot^\nu u^\mu)
\Gamma^{(\frac{3}{2})+}_{3~\mu\nu}\gamma_5C$
is just the same as equation (30).
This implies that the spin-$\frac{1}{2}$ baryon states represented by wave
functions $\gamma^{(\frac{1}{2})+}_5\gamma_\bot^\mu u\Gamma_{2~\mu}C$ and
$\sigma_\bot^{\mu\nu}u\Gamma^{(\frac{1}{2})+}_{\mu\nu}\gamma_5C$ 
degenerate with the spin-$\frac{3}{2}$
states represented by $u^\mu\Gamma^{(\frac{3}{2})+}_{1~\mu} C$ and
$\gamma_5(\gamma_\bot^\nu u^\mu -\gamma_\bot^\mu u^\nu)
\Gamma^{(\frac{3}{2})+}_{3~\mu\nu}\gamma_5C$
respectively. Similarly, the spin-$\frac{3}{2}$ baryon states represented by
the components, 
$\gamma_5 (\gamma_\bot^\nu u^\mu + \gamma_\bot^\mu u^\nu)
\Gamma^{(\frac{3}{2})+}_{4~\mu\nu}\gamma_5C$
and
$[\sigma_\bot^{\lambda\nu}u^{\mu} - \frac{1}{2}(g_\bot^{\lambda\mu}u^\nu
- g_\bot^{\nu\mu}u^\lambda)]\Gamma^{(\frac{3}{2})+}_{\mu\nu\lambda}C$
will degenerate with the spin-$\frac{5}{2}$ baryon states represented by the 
components $u^{\mu\nu}\Gamma^{(\frac{5}{2})+}_{1~\mu\nu}\gamma_5C$ and
$\gamma_5(\gamma_\bot^\lambda u^{\nu\mu} - \gamma_\bot^\nu u^{\lambda\mu})
\Gamma^{(\frac{5}{2})+}_{3~\mu\nu\lambda}C$ respectively.
The case for baryons with negative parity is just the same.
Such a conclusion is the direct result of heavy quark limit, and can be
generalized to any given spin baryon states.
\par{~~~}
Generally, when $j$ is even, the wave function of 
spin-$(j+\frac{1}{2})$ baryons with positive parity can be written as 
$$\begin{array}{l}
\chi^{(j+\frac{1}{2})+}(P,q,k)=u^{\mu_1\mu_2\ldots \mu_j}
\Gamma^{(j+\frac{1}{2})+}_{1~\mu_1\mu_2\ldots\mu_j}\gamma_5C
+ \gamma_5u^{\mu_1\mu_2\ldots \mu_j}
\Gamma^{(j+\frac{1}{2})+}_{2~\mu_1\mu_2\ldots \mu_j}C\\
\hskip 0.3in
+ (\gamma_\bot^{\mu_1}u^{\mu_2\mu_3\ldots \mu_{j+1}}
+\gamma_\bot^{\mu_2}u^{\mu_1\mu_3\ldots \mu_{j+1}}+\ldots 
+\gamma_\bot^{\mu_{j+1}}u^{\mu_2\mu_3\ldots \mu_j})
\Gamma^{(j+\frac{1}{2})+}_{1~\mu_1\mu_2\ldots\mu_{j+1}}\gamma_5C \\
\hskip 0.3in
+(\gamma_\bot^{\mu_1}u^{\mu_2\mu_3\ldots \mu_{j+1}}
-\gamma_\bot^{\mu_2}u^{\mu_1\mu_3\ldots \mu_{j+1}})
\Gamma^{(j+\frac{1}{2})+}_{2~\mu_1\mu_2\ldots\mu_{j+1}}\gamma_5C \\
\hskip 0.3in
+\gamma_5(\gamma_\bot^{\mu_1}u^{\mu_2\mu_3\ldots \mu_{j+1}}
+\gamma_\bot^{\mu_2}u^{\mu_1\mu_3\ldots \mu_{j+1}}+\ldots 
+\gamma_\bot^{\mu_{j+1}}u^{\mu_2\mu_3\ldots \mu_j})
\Gamma^{(j+\frac{1}{2})+}_{3~\mu_1\mu_2\ldots\mu_{j+1}}C \\
\hskip 0.3in
+\gamma_5(\gamma_\bot^{\mu_1}u^{\mu_2\mu_3\ldots \mu_{j+1}}
-\gamma_\bot^{\mu_2}u^{\mu_1\mu_3\ldots \mu_{j+1}})
\Gamma^{(j+\frac{1}{2})+}_{4~\mu_1\mu_2\ldots\mu_{j+1}}C \\
\hskip 0.3in
+[\sigma_\bot^{\mu_1\mu_2}u^{\mu_3\mu_4\ldots \mu_{j+2}}
-\frac{1}{j+1}(g^{\mu_1\mu_3}
u^{\mu_2\mu_4\ldots \mu_{j+2}}+g^{\mu_1\mu_4}
u^{\mu_2\mu_3\mu_5\ldots \mu_{j+2}}\ldots 
+g^{\mu_1\mu_{j+2}}u^{\mu_2\mu_3\ldots \mu_{j+1}})\\
\hskip 0.3in
+\frac{1}{j+1}(g^{\mu_2\mu_3}u^{\mu_1\mu_4\ldots \mu_{j+2}}
+g^{\mu_2\mu_4}u^{\mu_1\mu_3\mu_5\ldots \mu_{j+2}}\ldots 
+g^{\mu_2\mu_{j+2}}u^{\mu_1\mu_3\ldots \mu_{j+1}})]
\Gamma^{(j+\frac{1}{2})+}_{\mu_1\mu_2\ldots\mu_{j+2}}\gamma_5C
\end{array}
\eqno(31)$$
The spin-$(j-\frac{1}{2})$ wave function with positive parity is written as
$$\begin{array}{l}
\chi^{(j-\frac{1}{2})+}(P,q,k)=u^{\mu_1\mu_2\ldots \mu_{j-1}}
\Gamma^{(j-\frac{1}{2})+}_{1~\mu_1\mu_2\ldots\mu_{j-1}}C
+ \gamma_5u^{\mu_1\mu_2\ldots \mu_{j-1}}
\Gamma^{(j-\frac{1}{2})+}_{2~\mu_1\mu_2\ldots \mu_{j-1}}\gamma_5C\\
\hskip 0.3in
+ (\gamma_\bot^{\mu_1}u^{\mu_2\mu_3\ldots \mu_j}
+\gamma_\bot^{\mu_2}u^{\mu_1\mu_3\ldots \mu_j}+\ldots 
+\gamma_\bot^{\mu_j}u^{\mu_2\mu_3\ldots \mu_{j-1}})
\Gamma^{(j-\frac{1}{2})+}_{1~\mu_1\mu_2\ldots\mu_j}C \\
\hskip 0.3in
+(\gamma_\bot^{\mu_1}u^{\mu_2\mu_3\ldots \mu_j}
-\gamma_\bot^{\mu_2}u^{\mu_1\mu_3\ldots \mu_j})
\Gamma^{(j-\frac{1}{2})+}_{2~\mu_1\mu_2\ldots\mu_j}C \\
\hskip 0.3in
+\gamma_5(\gamma_\bot^{\mu_1}u^{\mu_2\mu_3\ldots \mu_j}
+\gamma_\bot^{\mu_2}u^{\mu_1\mu_3\ldots \mu_j}+\ldots 
+\gamma_\bot^{\mu_j}u^{\mu_2\mu_3\ldots \mu_{j-1}})
\Gamma^{(j-\frac{1}{2})+}_{3~\mu_1\mu_2\ldots\mu_j}\gamma_5C \\
\hskip 0.3in
+\gamma_5(\gamma_\bot^{\mu_1}u^{\mu_2\mu_3\ldots \mu_j}
-\gamma_\bot^{\mu_2}u^{\mu_1\mu_3\ldots \mu_j})
\Gamma^{(j-\frac{1}{2})+}_{4~\mu_1\mu_2\ldots\mu_j}\gamma_5C \\
\hskip 0.3in
+[\sigma_\bot^{\mu_1\mu_2}u^{\mu_3\mu_4\ldots \mu_{j+1}}
-\frac{1}{j}(g^{\mu_1\mu_3}
u^{\mu_2\mu_4\ldots \mu_{j+1}}+g^{\mu_1\mu_4}
u^{\mu_2\mu_3\mu_5\ldots \mu_{j+1}}\ldots 
+g^{\mu_1\mu_{j+1}}u^{\mu_2\mu_3\ldots \mu_j})\\
\hskip 0.3in
+\frac{1}{j}(g^{\mu_2\mu_3}u^{\mu_1\mu_4\ldots \mu_{j+1}}
+g^{\mu_2\mu_4}u^{\mu_1\mu_3\mu_5\ldots \mu_{j+1}}\ldots 
+g^{\mu_2\mu_{j+1}}u^{\mu_1\mu_3\ldots \mu_j})]
\Gamma^{(j-\frac{1}{2})+}_{\mu_1\mu_2\ldots\mu_{j+1}}C
\end{array}
\eqno(32)$$
We obtain the spin-$(j+\frac{3}{2})$ wave function with positive parity by
replacing $j$ with $j+2$ in the expression (32).\\
In the heavy quark limit, the second, the third and the fourth components
of the wave function (31) are
reduced to zero. The components left are decoupled. Instituting  the wave
function of spin-$(j+\frac{1}{2})$, spin-$(j-\frac{1}{2})$ and
spin-$(j+\frac{3}{2})$ into equation (25), we find the following degenerate
doublets. 
\[
\left\{ \begin{array}{l}
u^{\mu_1\mu_2\ldots \mu_j}
\Gamma^{(j+\frac{1}{2})+}_{1~\mu_1\mu_2\ldots \mu_j}\gamma_5C ~, \\
~~\\
\gamma_5(\gamma_\bot^{\mu_1}u^{\mu_2\mu_3\ldots \mu_j}
+\gamma_\bot^{\mu_2}u^{\mu_1\mu_3\ldots \mu_j}+\ldots
+\gamma_\bot^{\mu_j}u^{\mu_1\mu_2\ldots \mu_{j-1}})
\Gamma^{(j-\frac{1}{2})+}_{3~\mu_1\mu_2\ldots \mu_{j-1}}\gamma_5C ~;\hskip 1.5in
\end{array}
\right.
\]
$$
\left\{\begin{array}{l}
\gamma_5(\gamma_\bot^{\mu_1}
u^{\mu_2\mu_3\ldots \mu_{j+1}}
-\gamma_\bot^{\mu_2}u^{\mu_1\mu_3\ldots \mu_{j+1}})
\Gamma^{(j+\frac{1}{2})+}_{4~\mu_1\mu_2\ldots\mu_{j+1}}C ~,\\
~~\\~
[\sigma_\bot^{\mu_1\mu_2}u^{\mu_3\mu_4\ldots\mu_{j+1}}
-\frac{1}{j}(g^{\mu_1\mu_3}
u^{\mu_2\mu_4\ldots \mu_{j+1}}
+g^{\mu_1\mu4}u^{\mu_2\mu_3\mu_5\ldots \mu_{j+1}}
+\ldots + g^{\mu_1\mu_{j+1}}u^{\mu_2\mu_3\ldots \mu_j}) \\
+\frac{1}{j}(g^{\mu_2\mu_3}
u^{\mu_1\mu_4\ldots \mu_{j+1}}+g^{\mu_2\mu_4}
u^{\mu_1\mu_3\mu_5\ldots \mu_{j+1}}
+\ldots + g^{\mu_2\mu_{j+1}}u^{\mu_1\mu_3\ldots \mu_j})]
\Gamma^{(j-\frac{1}{2})+}_{\mu_1\mu_2\ldots \mu_{j+1}}C. ~;\hskip 1in
\end{array}
\right.
$$
\[
\left\{\begin{array}{l}
\gamma_5(\gamma_\bot^{\mu_1}
u^{\mu_2\mu_3\ldots \mu_{j+1}}
+\gamma_\bot^{\mu_2}u^{\mu_1\mu_3\ldots \mu_{j+1}}+\ldots 
+\gamma_\bot^{\mu_{j+1}}u^{\mu_1\mu_2\ldots \mu_j})
\Gamma^{(j+\frac{1}{2})+}_{3~\mu_1\mu_2\ldots\mu_{j+1}}C ~,\hskip 1in\\
~~\\
u^{\mu_1\mu_2\ldots \mu_{j+1}}
\Gamma^{(j+\frac{3}{2})+}_{\mu_1\mu_2\ldots \mu_{j+1}} ~;  
\end{array}
\right.
\]
\[
\left\{\begin{array}{l}
[\sigma_\bot^{\mu_1\mu_2}u^{\mu_3\mu_4\ldots\mu_{j+2}}
-\frac{1}{j+1}(g^{\mu_1\mu_3}u^{\mu_2\mu_4\ldots \mu_{j+2}}
+g^{\mu_1\mu_4}u^{\mu_2\mu_3\mu_5\ldots \mu_{j+2}}+\ldots 
+g^{\mu_1\mu_{j+2}}u^{\mu_2\mu_3\ldots \mu_{j+1}}) \\
+\frac{1}{j+1}(g^{\mu_2\mu_3}u^{\mu_1\mu_4\ldots \mu_{j+2}}
+g^{\mu_2\mu_4}u^{\mu_1\mu_3\mu_5\ldots \mu_{j+2}}+\ldots 
+g^{\mu_2\mu_{j+2}}u^{\mu_1\mu_3\ldots \mu_{j+1}})]
\Gamma^{(j+\frac{1}{2})+}_{\mu_1\mu_2\ldots \mu_{j+2}}\gamma_5C ~,\\
~~\\
\gamma_5(\gamma_\bot^{\mu_1}
u^{\mu_2\mu_3\ldots \mu_{j+2}}
-\gamma_\bot^{\mu_2}u^{\mu_1\mu_3\ldots \mu_{j+2}})
\Gamma^{(j+\frac{3}{2})+}_{4~\mu_1\mu_2\ldots \mu_{j+2}}\gamma_5C ~.
\hskip 2.2in
\end{array}
\right.
\]
The corresponding wave function  with negative parity have the
similar degeneracy.
\par
4.{\it Conclusion}~~~We construct the B.S equation for baryons. In this
framework we discuss the spin symmetry of heavy baryon 
system. The most general form of the B.S. wave function of a baryon is
obtained.  It is complicated, and the B.S. equations are
some coupled integral equations. 
In the heavy quark limit, the freedom of the heavy quark spin is
decoupled from the light ones, and consequently the spin symmetry takes place.
This
symmetry results in some degeneracy in the spectrum of heavy baryons. We
verified this point clearly in the framework of B.S. equation. 
Meanwhile, in the heavy quark limit, some components of the B.S. wave function
becomes to zero, and some others decoupled. Therefore, the B.S. 
equations of baryons are simplified greatly. 

ACKNOWLEGEMENT\\
  This work is supported in part by China National Natural Science Fundation.

\end{document}